# A New Theory
# ON THE MOTION OF WATERS
# THROUGH CHANNELS OF ANY KIND

By

Daniel Bernoulli, son of Johann

In the month of June 1727

Translated an Annotated by

Sylvio R. Bistafa

sylvio.bistafa@gmail.com

June 2025

Summary

This paper presents an annotated English translation of Daniel Bernoulli's 1727 work "A New Theory on the Motion of Waters through Channels of Any Kind", originally published in the *Commentarii Academiae Scientiarum Imperialis Petropolitanae*. Written over a decade before his renowned Hydrodynamica (1738), this early treatise reveals Bernoulli's foundational approach to fluid motion based on the conservation of vis viva—the kinetic energy of moving bodies—at a time when the principle was still under active debate. In this work, Bernoulli applies mechanical reasoning to fluids flowing through channels of arbitrary shape, deriving relationships between velocity, cross-sectional area, and efflux under ideal conditions. He anticipates core results of Hydrodynamica, including the inverse relationship between flow velocity and cross-sectional area, and emphasizes the role of energy balance in analyzing steady flow. The text also includes reflections on experimental limitations, the influence of friction, and the boundaries of theoretical applicability. This translation highlights the historical and conceptual significance of the 1727 paper as a precursor to Bernoulli's mature hydrodynamic theory.

Foreword

The text presented in this paper—Daniel Bernoulli's 1727 New Theory on the Motion of Waters through Channels of Any Kind, originally published in the Commentarii Academiae Scientiarum Imperialis Petropolitanae—offers a significant glimpse into the theoretical developments that would culminate in Bernoulli's seminal treatise Hydrodynamica (1738). In this earlier work, Bernoulli grapples with the complex behavior of water flowing through tubes of various shapes and orientations, combining physical intuition with mathematical rigor in a way that foreshadows many of the central arguments of Hydrodynamica.

---

[a] A copy can be found at the end of the translation.

Of particular importance in this 1727 contribution is the prominent use of the principle of conservation of vis viva—a precursor to what we now understand as kinetic energy. Bernoulli defines vis viva as the product of mass and the square of velocity, and uses it as a fundamental tool for analyzing fluid motion. While today the use of energy conservation principles is taken for granted in mechanics, at the time of this paper's publication the concept of vis viva was still controversial. It had been promoted by Leibnizians and further clarified by Huygens, but remained the subject of debate among Newtonians and other contemporary mathematicians, particularly with regard to collisions and fluid interactions.

D. Bernoulli's insistence on the validity of vis viva—and his demonstration of its applicability to fluid systems—marks a bold and visionary move. He explicitly acknowledges the resistance to this concept among his peers, even cautioning that "the mere mention of the conservation of living forces causes discomfort to some," but asserts its equivalence to the Huygenian principle of gravitational energy balance. By grounding his entire theory of fluid motion in the conservation of vis viva, Bernoulli anticipates the eventual unification of mechanics and hydrodynamics through energetic principles, decades ahead of the formal development of energy conservation laws.

The manuscript also establishes a foundational result that would echo through Hydrodynamica: that velocities of effluent water are inversely proportional to the cross-sectional areas of the channels, an idea that complements the conservation of vis viva and leads to a powerful predictive framework for analyzing flow. This principle would become central to Bernoulli's later derivation of pressure-velocity relationships in moving fluids.

Finally, this 1727 paper shows an early concern for the limits of theory. Bernoulli acknowledges that in irregular or discontinuous geometries—such as vessels with attached bags—the assumptions of his model break down. This recognition reveals a deep understanding not only of idealized mathematical systems but also of the empirical boundaries within which they operate.

In presenting this translation and commentary, the goal is to highlight how Bernoulli's early work laid the conceptual and mathematical groundwork for Hydrodynamica, and how it reflects both the philosophical tensions and scientific innovations of early 18th-century mechanics. This paper is not merely a historical curiosity but a vital piece in understanding the evolution of hydrodynamic thought and the long path toward the modern science of fluid motion.

_____________________

Many geometers, and very famous ones at that, have attempted to determine the motion of water through pipes; but few have provided anything that agrees with experiment, and none have established a complete theory. That water standing in a pipe will exit through a very small hole with such velocity as would allow it to rise to the height of the water's surface above the hole has been correctly determined by some mathematicians. But those who have sought to go further have produced nothing but conjectures—conjectures completely contrary to experiment.

I myself, after often understanding from my father the great utility of the principle of conservation of living forces (vis viva) for solving countless physico-mathematical problems—which would otherwise be considered very difficult, if not altogether hopeless—was struck with the idea that this same principle might serve for discovering the much-desired theory of flowing water in pipes. And the outcome did not deceive my expectation.

But so that I may now approach the matter itself more closely, I shall first say what is to be understood by living forces and their perpetual conservation.

Living force (vis viva) is said to be that which exists in a moving body, and it is measured by the product of the square of the velocity and the mass of the body. If several bodies are in motion, the total force, or quantity of forces, is to be estimated by summing all such defined products.

Huygens, and after him many others, demonstrated that this sum remains constant, however the bodies may collide with one another—provided they are perfectly elastic and conceived to be moving in a vacuum.

Thus, living forces are also conserved in elastic bodies. Now, I assume that the smallest particles making up a fluid are perfectly elastic—since if they were not extremely hard and endowed with maximum elasticity, they could be further subdivided.

With these preliminaries laid out, it is now clear what must happen when water flows horizontally through a pipe that is not cylindrical, but, for example, conical, since no particle of the water can continue its motion in a straight line, the particles strike the walls of the pipe and are reflected, and then in turn collide with other parts of the fluid. Meanwhile, during this agitation, the same quantity of living forces (vis viva) is perpetually conserved, and under this law the water will continue its motion through the pipe.

It must, however, be well noted that all the motions just described are minimal, so that no particle changes its position except to the extent that the entire mass of fluid has a progressive motion. This is just as we see in the case of many elastic balls placed next to each other in a straight line and of equal size: if the one at the end is struck, it is not the whole series of balls that moves, but only the one at the opposite end.

And in this way, it is not difficult for anyone to see that the same quantity of living forces may be conserved in the motion of fluids, exactly as in the motion of elastic bodies colliding with one another. Indeed, this must necessarily occur because of the supreme elasticity that is inherent in the minute particles of fluids.

Now I would proceed to the matter itself, were it not that I am perfectly aware that even the mere mention of the conservation of living forces causes discomfort to some. For their sake, I believe it proper to point out that this principle of conservation of living forces differs in no way from the principle first employed by Huygens and later accepted by all geometers without any controversy—namely, that bodies impelled downward by the force of gravity acquire such a velocity that if each were to ascend again with its final velocity, it would return to the same common center of gravity and come to rest it would return to its original height. Whoever prefers this Huygenian principle will resolve the matter with equal ease.

To this, however, we must add another principle: the velocities of fluids flowing through a pipe of varying cross-section are everywhere inversely proportional to the cross-sectional areas.

With these two principles, we shall resolve the entire problem.

Proposition I — Problem

Given the velocity with which the surface of water moves in any pipe, to find the living force (vis viva) of the entire mass of water.

Solution.

Let (Fig. I) ABGH be a vessel through which the liquid CDFE flows, whose current position is approximately cdfe. Let the surface CD have a velocity equal to that which a heavy body would acquire by falling from the height NO. It is clear that the velocity at LM stands to the velocity at the surface CD in the inverse ratio of the cross-sectional areas at CD and LM. Therefore, if the whole fluid is imagined as being divided into infinitely thin horizontal layers of the same width—like the layer at LMml—the vis viva of each layer is as its mass times the square of its velocity; that is, proportional to

$$\frac{LM}{CD} \times \frac{CD^2}{LM^2}$$

and since the velocities are inversely proportional to the cross-sectional areas, the living forces are inversely proportional to the cross-sectional areas, or as $\frac{CD}{LM}$.

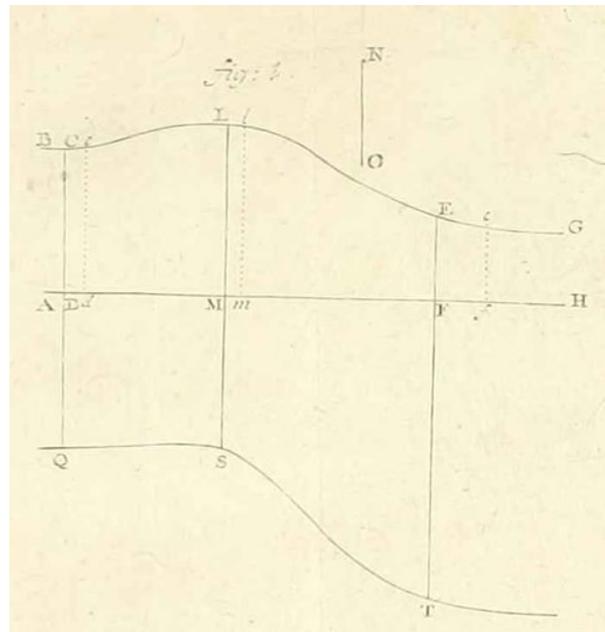

Hence it is understood that, if another curve QST is drawn along the same axis AH such that MS is everywhere equal to the "third proportional" (*tertiae continue*

*proportionali*)[b] to LM and CD, then the total living force of the entire mass of water CDFE will equal the space DQTF × NO.

If we prefer to use symbols, we will have the desired living force as:

$$aav \int \frac{dt}{s},$$

where $a$ denotes the area of surface CD, and $v$ the height through which a heavy body would have to fall to acquire that velocity, $dt$ denotes an element of flow and $s$ [denotes] the cross-sectional area of the vessel at ML.

Proposition II — Theorem

If the tube (fig. 1) ABGH is placed vertically, and a mass of water descends under its own weight into the position CDFE, which it then immediately changes into the position cdfe, then the differential of the vis viva — that is, the increment of vis viva acquired during that moment — is equal to that which would be acquired by a cylindrical mass of water whose base is CD and whose height is DF, falling freely.

Demonstration

The vis viva acquired must be calculated from the mass and the height of the descent. Now, when the volume CDFE arrives at the position cdfe, it is the same as if the water cdFE had remained in place, and the volume CDdc had arrived at the position EF. Therefore, the newly acquired vis viva is: $CD \times dD \times DF$ or equivalently, $CD \times DF \times Dd$. Q.E.D. (*Quod Erat Demonstrandum* – Which was to be demonstrated).

Proposition 3 — Problem

To determine the velocity of the water which flows at each instant through a tube of any shape and perforated by an opening of any kind.

Solution.

Let BDFG (fig. 2) be any curve whose horizontal ordinates DC represent, respectively, the cross-sectional areas of the tube. Let the water descend from point A to C, so that: $AC = t$, $CD = s$, the cross-sectional area of the opening represented by LM is $b$. Let the total height $AM = c$. The total vis viva (kinetic energy) contained in the fluid when it is in the position DCMG is $M$. But the velocity of the surface at DC is equal to that which a body acquires by falling from height $v$. Now imagine that the water from

---

[b] The Latin phrase *tertiae continue proportionali* means "the third in continued proportion", or more precisely, "the third proportional in a continued proportion". In mathematical terms, if you have two quantities $a$ and $b$, the third proportional $c$ is defined by the proportion: $\frac{a}{b} = \frac{b}{c}$. Solving for $c$, you get: $c = \frac{b^2}{a}$. This concept goes back to classical Greek mathematics and was common in early modern mathematical Latin. It's essentially saying that: a, b, and c are in continued proportion if: $\frac{a}{b} = \frac{b}{c}$, then $c$ is the third proportional to a and b. In the Bernoulli context: Bernoulli refers to constructing a curve such that a length (called MS) is everywhere equal to the third proportional to two other segment lengths (LM and CD). In practice, this means: $MS = \frac{CD^2}{LM}$. This is used to geometrically represent how the kinetic energy (vis viva) of fluid layers varies with cross-sectional area along a pipe.

the position DCMLGD has reached the position FEONLGF. Then, by Proposition II, the differential of vis viva is: $DC \times CM \times CE = s \times (c - t) \times dt$. But the same increment can also be defined in another way: while the water surface was at DC, the velocity at FE was $\frac{s}{s+ds}\sqrt{v}$ [c](for the velocities are in the inverse ratio of the cross-sectional areas), and the vis viva of the water FEMLGF was $M - svdt$. Now, after the surface of the water has passed from point DC to FE, it has a velocity of $\sqrt{v} + \frac{dv}{2\sqrt{v}}$; and since vis viva is in the duplicate (i.e., squared) ratio of the velocities, the vis viva of the water FEMLGF will be:

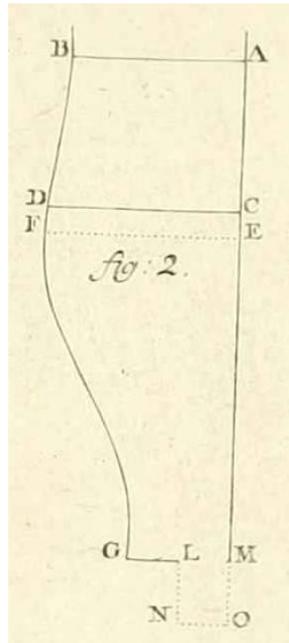

$$(M - svdt) \times \left(\sqrt{v} + \frac{dv}{2\sqrt{v}}\right)^2 \div \left(\frac{s}{s+ds}\sqrt{v}\right)^2$$
$$= \text{(having neglected what ought to be neglected)} \frac{(s+ds)^2(Mv + Mdv - sv^2dt)}{s^2v}$$
$$= \frac{(s^2 + 2sds)(Mv + Mdv - sv^2dt)}{s^2v} = \frac{Msv - s^2v^2dt + +Msdv + 2Mvds}{sv}.$$

If to this quantity is added the vis viva of the droplet LMON, which has just exited the tube, we obtain the total vis viva possessed by the water after the change of configuration. Now, the water particle LMON is equal to DCEF = $sdt$, and it has such velocity as would allow it to rise to a height of $\frac{s^2}{b^2}v$; from which its force (or energy) follows $\frac{s^3}{b^2}vdt$; therefore, the total vis viva of the water FEONLGF is:

$$\frac{Msv - s^2v^2dt + +Msdv + 2Mvds}{sv} + \frac{s^3}{b^2}vdt,$$

---

[c] Here, D. Bernoulli may have inadvertently used $u$ in place of the previously defined variable $v$, which he had explicitly introduced as a velocity head. He then proceeds to differentiate $u$ without ever defining it, and the resulting expressions behave as if $u = v$. This appears to be a notational or editorial lapse — a common occurrence in early mathematical treatises. In our interpretation, we have instead replaced $u$ with $v$ wherever $u$ appears inconsistently throughout the work.

Therefore, if $M$ is subtracted, the differential of the vis viva will be obtained, which corresponds to

$$\frac{Msdv + 2Mvds - s^2v^2dt}{sv} + \frac{s^3}{b^2}vdt.$$

From this, the following equation is obtained

$$s(c-t)dt = \frac{Msdv + 2Mvds - s^2v^2dt}{sv} + \frac{s^3}{b^2}vdt.$$

Moreover, by Proposition 2, the quantity $M$ can be obtained, and $s$ is given in terms of $t$; thus, an equation between $t$ and $v$ is obtained, by means of which the velocity of the water can be determined at any position. Q.E.I. (*Quod Erat Inveniendum* – Which was to be determined).

Proposition 4 — Problem
To determine the velocities with which water flows at each instant from a vertical cylindrical tube, whether the orifice is finite or infinitely small.

Solution.

Let the cross-sectional area of the cylinder (perpendicular to the axis) be $= n$, the cross-sectional area of the orifice be $= 1$, the total height of the water column before efflux be $= c$, and the vertical distance that the upper surface of the water has descended be $= t$. Then, the remaining vis viva (kinetic energy) of the water in the tube — which we earlier denoted by $M$ — is $= n(c-t)v$. What we previously referred to generally as $s$, is now taken to be constant and equal to $n$. Substituting these values into the canonical equation, we obtain:

$$n(c-t)dt = n(c-t)dv - nvdt + n^3vdt;$$

Let it be $(c-t) = z$ and $n^2 - 1 = m$ and there will arise

$$-zdz = zdv - mvdz.$$

This equation, reduced to logarithmic differentials by the "paternal method" (i.e., the calculus methods introduced by his father Johann Bernoulli)[d], and properly worked out, gives:

---

[d] The phrase "reduced to logarithmic differentials by the paternal method" reflects an early integration technique commonly used by Johann Bernoulli and his contemporaries. The trick is that equations of the form: $\frac{dv}{v} = (function) \cdot dz$ or $\frac{dv}{v} = (function) \cdot dz$, lead to logarithmic integration $\int \frac{dv}{v} = \ln v + C$ or $\int \frac{dz}{z} = \ln z + C$. So, the term "logarithmic

$$v = \frac{c^{n^2-2}z - z^{n^2-1}}{(n^2-2)c^{n^2-2}}.^{\text{e}}$$

Q.E.I. (Quod Erat Inveniendum – "which was to be found")

Corollary 1.

If $n = 1$, that is, if there is no bottom in the tube (i.e., the fluid falls freely through the entire height $c$), then the velocity of the water is the same as that which a heavy body would naturally acquire by falling through the height $c - z$, which any person, even without setting up a formal calculation, could have determined[f]. Nevertheless, there have been those who even in this case believed that water immediately from the start flows out with the same velocity that a body acquires by falling through the entire height of the water column.

Corollary 2.

There is always a point in the tube at which, if the surface of the water reaches it, the velocity of the efflux becomes maximum; this point is reached when one takes

$$z = \frac{c}{(n^2-1)^{1/(n^2-2)}},$$

such that[g]

$$v_{\max} = \frac{c^{2-n^2} \cdot \left(c^{2-n^2}(n^2-1)\right)^{-1/(n^2-2)} \left[c^{n^2-2} - \left(c^{2-n^2}(n^2-1)\right)^{1/(n^2-2)} \cdot \left(c^{2-n^2}(n^2-1)\right)^{-(n^2-1)/(n^2-2)}\right]}{n^2-2}$$

This quantity gives the maximum velocity that the surface of the water in the tube can reach in descending; and if we multiply this same quantity by $n^2$, we obtain the height needed to generate the maximum velocity of the effluent water, which is always less than $c$. But if $n$ becomes infinite, this height degenerates into $c$ itself, from which it is also manifest by our method that in the case of an infinitely small orifice, the water exits with the same velocity as a body would acquire in falling from the height $c$; and this is the only case which writers on hydraulics have correctly understood.

---

differentials" indicates that after algebraic manipulation, the equation becomes separable and integrable using natural logarithms.

[e] $-zdz = zdv - mvdz$ can be rewritten as $z\frac{dv}{dz} = mv - z$, and its solution can be found in the Appendix.

[f] For $n = 1$, which implies $m = n^2 - 1 = 0$, the differential equation becomes: $z\frac{dv}{dz} = -z$ or: $\frac{dv}{dz} = -1$. Integrating gives: $v(z) = -z + const$. Applying the boundary condition $v(c) = 0$, we find const$= c$ and $v(z) = c - z$.

[g] The corresponding expression printed in the original manuscript is quite difficult to read, so we instead generated it using ChatGPT. Although the resulting expression is exact, it is rather intricate due to the involved exponent manipulations.

But if $n$ is not infinite, but still sufficiently large, then the maximum efflux velocity of the water will not be much less than that which would occur if the orifice were infinitely small. For instance, if $n = 10$, the water flowing out at its maximum velocity could rise to about $\frac{97}{100}$ of the height $c$ [h]; and this maximum velocity is almost immediately acquired at the beginning of the flow, namely after the water has descended in the tube through a distance of $\frac{47}{1047} c$ [i]. All of this agrees excellently with the experiments.

Corollary 3.

If $n$ is a large number, and the water has already begun to descend somewhat in the tube, then the velocities will be approximately as the square roots of the water heights. This rule has been assumed by those who have written on the division of clepsydrae (water clocks), such as the esteemed Varignon and Mariotte. I myself, when once dealing with the use of a spherical clepsydra, considered the matter in the same way. However, the rule fails somewhat at the very beginning of the efflux, unless $n$ is a very large number. For a true (accurate) division, it is required that one perform the integration of $\frac{dz}{\sqrt{v}}$ which, however, can never be done in all cases[j]; nonetheless, series can be developed for this purpose, by which the total time of evacuation may be determined approximately in each particular case. This gave me the occasion to carry out numerous experiments using various tubes and different orifices; and in every case, the time of depletion as measured by experiment agreed precisely with that predicted by calculation. Thus, our theory can now be considered beyond doubt. Moreover, I have additional experiments at hand by which the matter can be subjected again to no less certain scrutiny.

Corollary 4.

---

[h] According to the above given expression, for $n = 10$, the maximum value of $v(z)$ is approximately: $v_{max} \approx 0.00964 \cdot c$, meaning that the maximum height from which a body would have to fall to gain the same velocity as the water exiting the tube is only about 0.964% (less than 1%) of the total height $c$, a proof that the expression given by ChatGPT is correct.

[i] The numerical value $\frac{47}{1047} \approx 0.0449$ or 4.49% of the total height $c$ does not match Bernoulli's own approximation for $n = 10$, which was: "...after the water has descended in the tube through a distance of $\frac{1}{104} c \approx 0.0096c$ ($\approx 0.96\%$). So, to align with Bernoulli's original statement, the sentence should read: "and this maximum velocity is almost immediately acquired at the beginning of the flow, namely after the water has descended in the tube through a distance of $\frac{1}{104} c$ "

[j] Bernoulli uses this differential expression in passages where he integrates over the vertical extent of a water column to calculate the total depletion time — e.g., how long it takes for a clepsydra to empty. By integrating $\frac{dz}{\sqrt{v(z)}}$ from the top of the water level to the orifice, he accounts for the varying speed of efflux as the pressure head decreases. $\frac{dz}{\sqrt{v(z)}}$ comes from the fundamental relationship: $time = distance/speed$ where the speed is interpreted via energy conservation as $\sqrt{2gv}$, with $v$ being the potential energy height.

The authors cited have given a curve for the clepsydra that produces a uniform descent — that is, one in which the fluid surface descends uniformly over time. But that curve is inaccurate both at the beginning and at the end, and is only approximately correct in the middle — and only when the orifice is very small. The true curve is deduced from our canonical equation in Proposition 3, by assuming $v$ to be constant and eliminating the term containing $dv$. However, this curve is far from being algebraic; in fact, if one wishes to eliminate the quantity $M$ and substitute its equivalent, the result involves second-order differentials. Nonetheless, it is notable that the problem of equable efflux can be made algebraic: namely, if (see fig. 3) a curve AND is sought whose rotation about the axis BF generates a solid such that, if fluid descends within it — as in a clepsydra — equal volumes of water are discharged in equal time intervals, or equivalently, that the fluid flows out with constant velocity. Then, letting $FM = t$, $MN = y$, $e$ equal the height FB, and $l$ is the height from which a falling body would acquire the efflux velocity, the curve of equable efflux is expressed by the equation: $y^4 \cdot t - y^4 \cdot l + c^4 \cdot l = 0$. This curve possesses many properties, the enumeration of which I omit so as not to be overly lengthy; but what I think should by no means be left unsaid is that this figure for the vessel of equable efflux is the same as the English 'cataract', which they used in an attempt to explain the phenomena of effluent water. Moreover, to further confirm our theory, I shall briefly speak of cylindrical vessels to which are attached other narrower cylindrical tubes, either placed vertically (whose phenomena were recently publicly presented by the celebrated Bulffinger), or horizontally, whether abruptly ending or indefinitely long. The consideration of the latter is not unimportant in physiological matters, as I shall show on another occasion.

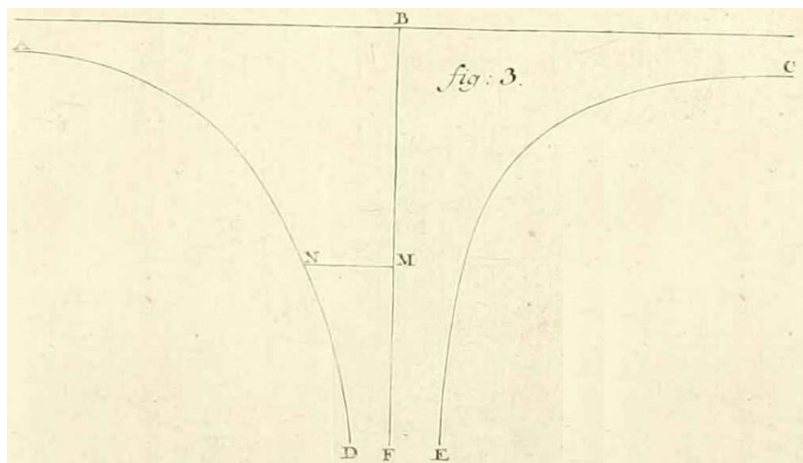

Proposition 5 [k]. Problem.

Given water flowing through the cylinder ACDB (fig. 4), into which a cylindrical tube EF is inserted, determine the velocity everywhere with which the water surface GH descends.

Solution.

---

[k] This was mistakenly written as Proposition 4 in the original publication.

This case can be regarded as a corollary of Proposition 3, and thus the velocity of the water surface can be determined throughout the entire process until all the water has flowed out of the vessel ACDB. The motion through the remaining part, namely the tube EF, is easily determined by itself, since it follows the laws of uniformly accelerated motion. Now, if the vessel was initially filled up to AB, and if we denote: $AG = t, AC = e, EF = c$, the surface $GH = n$, the cross-sectional area (amplitude) of the tube $EF = 1$, the height necessary to generate the velocity of the surface $GH = v$, then the canonical equation, with all appropriate substitutions made, will be transformed into a form suited to this particular case. Letting $(c + e - t) \times dt = (e - t + nc) \times dv + (n^2 - v) \times dt$. The same equation can thus be derived in this particular way. The vis viva (living force) of the entire mass of flowing water GCEFEDH is: $n \times (e - t) \times v + n^2 cv$, whose differential is: $(ne - nt + n^2 c)dv - nvdt$. If we add the vis viva of the particle FP that has flowed out, which is: $n^3 vdt$, then the total increment of vis viva generated in that moment is obtained, which, by Prop. 2, is equal to: $(e - t + c)ndt$. And dividing both sides by $n$, we again arrive at: $(c + e - t) \times dt = (e - t + nc) \times dv + (n^2 - v) \times dt$. Between these two methods, the difference is that by the first, the matter could be treated much more generally than by the second, which assumes neither the vessel nor the tube to be cylindrical, but of another shape. To reduce the latter equation to finite quantities, let: $t = q + e + nc, and\ v = r - \frac{c}{n+1}$, and then it transforms into this: $-qdq = -qdr + (n^2 - 1)rdq$, which, with the appropriate constant added, passes over into this: $q^{1-n^2} \times (2r - n^2 r - q) = (-n - nc)^{1-n^2} \times \left(\frac{ne+e+nc+2c}{n+1}\right)$ or $\left(\frac{n^2 r - 2r + q}{q^{n^2-1}}\right) = (nc + 2c + ne + e) \div (-n - 1) \times (-e - nc)^{n^2-1}$. And, with the quantities $t$ and $v$ substituted back in, we obtain: $(t - e - nc)^{1-n^2} \times \left(-n^2 v + 2v + c - t + e + \frac{c}{n+1}\right) = (-e - nc)^{1-n^2} \times \left(\frac{ne+e+nc+2c}{n+1}\right)$, or finally: $v = \left(e + c + \frac{c}{n+1} - t\right) \div (n^2 - 2) - \left(e + c + \frac{c}{n+1}\right) \times (t - e - nc)^{n^2-1} \div (n^2 - 2) \times (-e - nc)^{n^2-1}$.

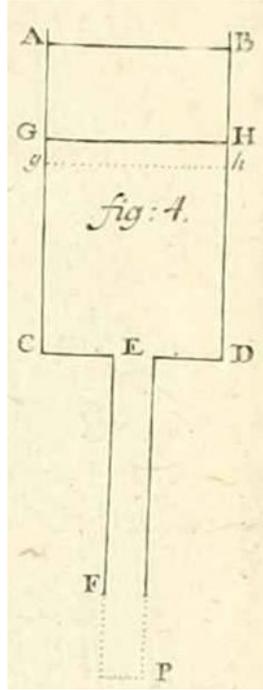

Corollary I. If we set *c = 0*, the theorem of Proposition 4 arises.[l]

Corollary II. If the maximum velocity is compared with the maximum velocity of water efflux in the case where $c = 0$, it will be found that the former is much greater than the latter; and indeed, the difference is the greater, the longer the tube EF is. Hence, it is not surprising that the times of emptying are the shorter, the longer the attached tubes are. However, all these things can be more precisely subjected to calculation, both in terms of velocities and in terms of times.

Corollary III. If the tube EF is indefinitely long, another reasoning — though not very different — will yield the following equation $edv - tdv + n^2 vdt + n^2 tdv - vdt = edt - tdt + ntdt$.[m]

---

[l] For *c = 0*, the expression becomes $v = \frac{(e-t)}{(n^2-2)} - \frac{e(t-e)^{n^2-1}}{(n^2-2)(-e)^{n^2-1}}$. Meanwhile, according to Proposition 4, the velocity $v$ is given by: $v = \frac{c^{n^2-2}z - z^{n^2-1}}{(n^2-2)c^{n^2-2}}$, which, after standardizing the notation (setting $z = e - t$), becomes: $v = \frac{e^{n^2-2}(e-t) - (e-t)^{n^2-1}}{(n^2-2)e^{n^2-2}}$. The two expressions for $v$ are mathematically equivalent. When we compute the difference between them, it simplifies to zero after accounting for the fact that: $(-e + t)^{n^2} = -(e - t)^{n^2}$ when $n^2$ is odd. This means that $\frac{(e-t)}{(n^2-2)} - \frac{e(t-e)^{n^2-1}}{(n^2-2)(-e)^{n^2-1}} = \frac{e^{n^2-2}(e-t) - (e-t)^{n^2-1}}{(n^2-2)e^{n^2-2}}$. Thus, both forms express the same function for $v$.

[m] This expression appears in the context of D. Bernoulli's analysis of unsteady outflow through a tube and seems to derive from the conservation of vis viva (living force), i.e., an energy balance principle, under the assumption that the tube EF is indefinitely long. In Proposition 4, D. Bernoulli analyzes the flow of water from a vessel with a cylindrical tube attached (EF). The general approach is to: compute the total differential of the vis viva (kinetic energy) of the fluid

Corollary IV. If the tube EF is slightly wider at point F than at point E, the time of depletion will be even shorter; indeed, the width at F can be increased to the point where the water, during efflux, ceases to adhere to the sides of the tube.

Corollary V. If the tube EF is inclined, with a few changes the calculation remains the same; therefore, I will not address these cases, except for the one in which the tube is indefinitely long and horizontal, since that case may be useful on other occasions. So, if the cylinder ACDB (fig. 5), filled with water up to AB, has an attached horizontal tube DE that is indefinitely long, and if the water in the cylinder has descended to MN while in the tube it has progressed to point R, let AC be called $= a$, the cross-sectional area AB $= n$, the orifice $= 1$, $CM = z$, and the speed of the water at MN such as would be acquired by falling from height $v$. I say that the equation between $z$ and $v$ will be as follows: $zv - n^2zv + n^2av - \frac{1}{2}z^2 + \frac{1}{2}a^2 = 0$ or $v = \frac{a^2 - z^2}{2(z - n^2z + n^2a)}$.

Scholium. I also carried out experiments with vessels to which narrower tubes were attached. But, as I had foreseen, the emptying times were always greater than what they should have been according to calculation. This must be attributed to friction: for when water issues through a simple orifice, there is hardly any friction; but when it flows through narrower channels, the sides of the tubes present resistance. I have a method for reducing such resistances to calculation through a single experiment— provided it is conducted with the utmost accuracy; for it is not difficult to see that these resistances are in inverse proportion to the diameters, and directly proportional to the velocities and lengths of the tubes. These resistances must be subtracted from the pressure of the water that drives the motion. To give an example of the method, let us consider a very large vessel filled with water, into which a narrow cylindrical tube is inserted horizontally. Let the height of the water above the tube be $= a$, the diameter of the tube $= b$, its length $= c$, and the velocity of the water exiting the tube such as would correspond to a fall from height $x$. Then the resistance of the tube will be: $= \frac{nc\sqrt{x}}{b}$ [n] (where $n$ is a number determined experimentally), which, when subtracted from the pressure of the water (which, assuming the orifice is very small compared to the vessel, is proportional to the height a of water above the tube), gives: $\frac{nc\sqrt{x}}{b} - a$, which quantity at the same time expresses the height to which the water

---

in the system (vessel + tube), equate it to the differential of the potential energy lost, derive an equation governing the velocity $v$ of the surface of the tube GH (which here is measured in height units). When EF is indefinitely long, certain simplifications and limits occur. Specifically: the volume and hence the kinetic energy associated with water in the tube EF dominate, the velocity at the vessel surface (GH) becomes negligible compared to that in the tube.

[n] This formula is dimensionally consistent and qualitatively correct for laminar flow, even though it lacks explicit use of viscosity. This suggests remarkable empirical insight — D. Bernoulli recognized frictional resistance as: increasing with pipe length, decreasing with pipe diameter, proportional to flow speed. All in line with modern low-Reynolds-number pipe flow.

could ascend with its own velocity; whence $\frac{nc\sqrt{x}}{b} - a = x$; and $x = \frac{n^2c^2 - 2ab^2 + nc\sqrt{(n^2c^2 - 4ab^2)}}{2b^2}$.[o]

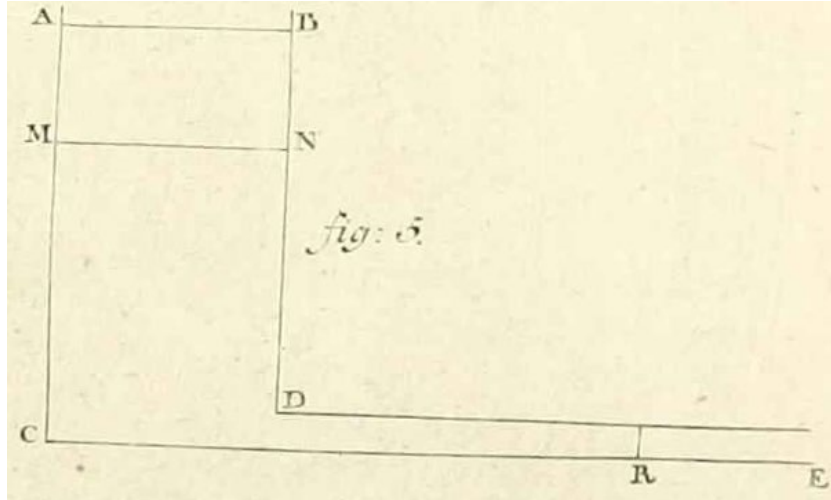

fig: 6.

Proposition 5. The problem can also be solved more elegantly in a general way, by the method of the third proposition. Let (fig. 6) BDG be a curve whose vertical axis is AM. Let the abscissae MC denote the residual water heights, and the ordinates CD express the width (amplitude) of the vessel at that location, that is, of the water surface. Let GM represent the bottom of the vessel and LM the orifice. Now let another curve SRP be constructed on the same axis, such that everywhere CR is the third proportional between DC and LM; and let: $MC = t$; $CD = s$; $GM = c$; $LM = b$, the velocity of the effluent water LNOM be such as would be acquired by the free fall of a body from height $v$, so $CR = \frac{b^2}{s}$, and (by Prop. I) the live force (vis viva) contained in the fluid DCMG is proportional to the space CRFM×$v$. Now suppose the space CRPM = N, and consider that as MC decreases, so too does N, $t$ and $v$. Then the increment of the live force of the water remaining in the vessel is $= -Ndv - vdN$ or, by substituting for $-dN$ its value $-\frac{b^2 dt}{s}$, we get: $= -Ndv - \frac{b^2 v dt}{s}$. To this, add the live force of the effluent particle LNOM, which is $= -svdt$, there will arise a total increment of live force (vis viva), to be equated with $-svdt$, that is: $-Ndv - \frac{b^2 v dt}{s} = -svdt$, or $dv + \frac{b^2 + s^2}{N} v dt = \frac{st dt}{N}$. Now let $\frac{b^2 + s^2}{N} dt = dP$ and let $\frac{st dt}{N} = dQ$, then $dv + vdP = dQ$, and let there be $v = h^{-P}$ (understanding by $h$ the number whose logarithm is unity), therefore: $h^{-P} dR = dQ$ and $R = \int h^P dQ$, and finally: $v = h^{-P} \int h^P dQ$. Q.E.I.

---

[o] The expression $x = \frac{n^2c^2 - 2ab^2 + nc\sqrt{(n^2c^2 - 4ab^2)}}{2b^2}$ is the solution to $\frac{nc\sqrt{x}}{b} - a = x$, obtained by solving the resulting quadratic equation in $x$.

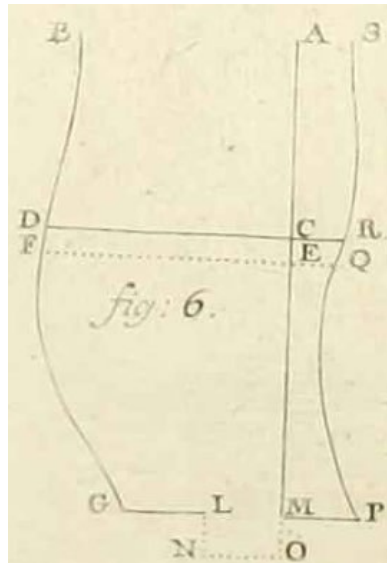

  There would be many more things to say — for instance, about elastic fluids, about fluids flowing out of a vessel through two orifices placed in any way whatsoever (this latter problem is extremely difficult), and other such matters; but let these suffice for now. Furthermore, as a corollary, the present theory also yields the doctrine of oscillating fluids in tubes — a topic that Newton touched upon in Proposition 35, Book II of the *Principia Mathematica Philosophiae Naturalis*, p. 363, 3rd edition. There, this oscillatory motion is correctly defined and is in agreement with our own principles, which are thus excellently confirmed. Indeed, it certainly suffers not the slightest objection, provided that two principles are maintained: that the conservation of live forces (vis viva), and that velocities are inversely proportional to the amplitudes be granted, in case frictions or other external resistances are not taken into account. The first principle cannot rightly be doubted unless one rejects all of mechanics entirely, since the conservation of forces (vis viva) has been accepted by all, merely under different terms. As for the second principle, although it may not be rigorously true, it can nevertheless be accepted without any hesitation in most vessels — namely, all those in which no sudden transition occurs. But, for example, if a vessel (Fig. 7) were taken as ABCD, perforated at E, and to which something like a bag were attached at O, then no one would fail to see that the motion of the water contained in the bag O would be quite different from what that principle would require; and therefore, the theory cannot be extended to such highly irregular vessels. I believe, however, that in such cases no certain rule can be established.

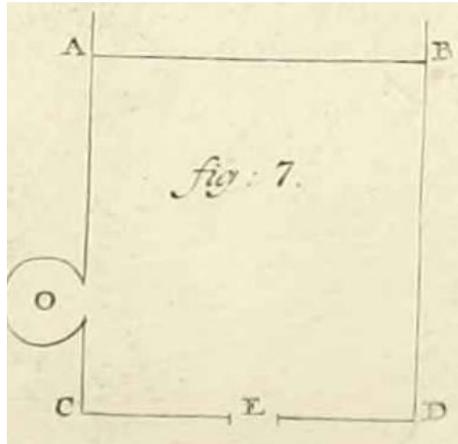

____________________

## Notes

The translation and some of the accompanying texts and comments were generated with the assistance of AI technology.

OpenAI, the organization behind ChatGPT, recommends providing appropriate context and verifying information against reliable sources, as AI-generated content is intended to support and augment human knowledge rather than serve as a definitive source.

The present interaction was conducted using GPT-4, with a knowledge cutoff in September 2023.

## Declarations

No funding was received for this study.

The author declares no competing interests relevant to the content of this article.

# APPENDIX

$$z\frac{dv}{dz} = mv - z$$

This is a **first-order linear ordinary differential equation (ODE)** in the unknown function $v(z)$, where $m$ is a constant.

## 🔄 Step 1: Standard Form

We rewrite the equation in standard linear ODE form:

$$\frac{dv}{dz} - \frac{m}{z}v = -1$$

This is of the form:

$$\frac{dv}{dz} + P(z)v = Q(z)$$

with

- $P(z) = -\frac{m}{z}$
- $Q(z) = -1$

## 📒 Step 2: Integrating Factor

We find the **integrating factor** $\mu(z)$, defined as:

$$\mu(z) = \exp\left(\int P(z)\,dz\right) = \exp\left(\int -\frac{m}{z}\,dz\right) = \exp(-m \ln z) = z^{-m}$$

## ✅ Step 3: Multiply Through by Integrating Factor

$$z^{-m}\frac{dv}{dz} - \frac{m}{z}z^{-m}v = -z^{-m} \Rightarrow \frac{d}{dz}\left(z^{-m}v\right) = -z^{-m}$$

## 📘 Step 4: Integrate Both Sides

Integrate both sides with respect to $z$:

$$\int \frac{d}{dz}(z^{-m}v)\,dz = \int -z^{-m}\,dz \Rightarrow z^{-m}v = \frac{z^{-m+1}}{m-1} + C \quad \text{(for } m \neq 1\text{)}$$

## 🔁 Step 5: Solve for $v(z)$

$$v(z) = z^m \left( \frac{z^{-m+1}}{m-1} + C \right) = \frac{z}{m-1} + Cz^m$$

Or rearranged:

$$v(z) = \frac{C(m-1)z^m + z}{m-1}$$

---

## ✳️ Step 6: Apply Initial/Boundary Condition (e.g., $v(c) = 0$)

Suppose we're given $v(c) = 0$, solve:

$$0 = \frac{C(m-1)c^m + c}{m-1} \Rightarrow C = -\frac{1}{c^{m-1}(m-1)}$$

Substitute back:

$$v(z) = \frac{-z^m + c^{m-1}z}{m-1} = \frac{c^{m-1}z - z^m}{m-1}$$

## 🎯 Final Result

$$v(z) = \frac{c^{m-1}z - z^m}{m-1} \quad \text{or, in terms of } n \text{ with } m = n^2 - 1: \quad v(z) = \frac{c^{n^2-2}z - z^{n^2-1}}{n^2 - 2}$$

## 🔁 Our derived result (from solving the ODE):

$$v(z) = \frac{c^{n^2-2}z - z^{n^2-1}}{n^2 - 2}$$

## 📜 Daniel Bernoulli's expression:

$$v(z) = \frac{c^{n^2-2}z - z^{n^2-1}}{(n^2 - 2) \cdot c^{n^2-2}}$$

## ✅ Resolution: Change of Variables

Bernoulli likely **non-dimensionalized** his equation by dividing lengths by $c$, the total height of the water column, setting $\xi = z/c$. Then:

$$v(\xi) = \frac{c \cdot (\xi - \xi^{n^2-1})}{n^2 - 2} \Rightarrow v(z) = \frac{z - \left(\frac{z}{c}\right)^{n^2-1} c}{n^2 - 2} = \frac{c^{n^2-2}z - z^{n^2-1}}{(n^2-2)c^{n^2-2}}$$

Exactly Bernoulli's formula.

This page intentionally left blank.

ad enodationem huiusmodi quaestionum iure requiri possunt, abunde hic exhibuisse mihi videor. Dedi enim primo generalissimas aequationes applicatu faciles: secundo methodum dedi infinitas aequationes vniuersales algebraicas inueniendi, ex quibus simplicissimas reipsa deduxi.

Tandem, quae ex transcendentalibus curuis cognitae sunt, etiam ex aequationibus generalibus facile deriuantur. Hisce omnibus praemisi solutionem duorum problematum agnatorum, de Pantogonia infinitorum axium traiectoria et de traiectoriis datum axium numerum habentibus, quippe quae ex consideratione naturae traiectoriarum reciprocarum sponte fluunt.

# *Theoria Noua* DE MOTV AQVARVM PER CANALES QVOSCVNQVE FLVENTIVM

## *Auctore* Daniele Bernoulli, Ioh. Fil.

Motum aquarum per tubos determinare aggressi sunt multi Geometrae iique celeberrimi; sed pauci aliquid dederunt, quod experientiae esset conforme, nemo autem integram theoriam stabiliuit. Aquam in tubo stagnantem

*M. Iun. 1727.*

per

per foramen valde paruum ea exilire velocitate, qua posfit afcendere ad altitudinem fuperficiei aqueae fupra foramen, a Mathematicis quibusdam recte fuit definitum; qui vlterius progredi voluerunt, nihil praeter coniecturas omni experientiae repugnantes protulerunt. Ego vero poftquam faepius intellexiffem ex Patre fummum vfum, quem habeat principium conferuationis virium viuarum pro infinitis problematis Phyfico-Mathematicis foluendis, quae alias pro valde difficilibus, ne dicam defperatis, habenda effent, mentem fubiit, num non idem principium pro eruenda theoria aquarum fluentium per tubos tantopere defiderata inferuire poffit, neque euentus fpem meam fefellit. Verum vt iam paulo propius ad rem ipfam accedam, dicam ante omnia, quid per vires viuas earumque perpetuam conferuationem intelligendum fit ; Dicitur itaque vis viua, quae ineft corpori moto atque menfuratur ex producto velocitatis quadrati in corporis maffam : fi plura corpora moueantur, vis totalis feu quantitas virium erit aeftimanda ex aggregato omnium productorum modo definitorum. Demonftrauit autem Hugenius poft eumque multi alii effe hoc aggregatum conftans quomodocunque fe inuicem percutiant ipfa corpora, modo fint perfecte elaftica atque in vacuo mota concipiantur. Conferuantur ergo etiam vires viuae in corporibus elafticis. Pono autem corpufcula minima fluidum aliquod componentia effe perfecte elaftica ; nifi enim effent duriffima fummaque elafticitate proinde praedita, poffent vlterius fubdiuidi. Hifce praemonitis manifeftum factum eft, quid fieri debeat, quando aqua fluit horizontaliter per tubum non cylindricum, sed v. gr. conicum ; nempe cum omnis aqua motum suum in linea recta continuare nequeat, particulae ipsius impingunt in latera tubi, et inde reflectunt, rursusque alias fluidi partes percutiunt ; interim durante hac agitatione eadem quantitas virium perpetuo conseruabitur hacque lege motum suum in tubo continuabit aqua. Notandum tamen probe est, praedictos motus omnes esse minimos, ita vt nulla particula locum suum mutet, nisi quatenus cum tota fluidi massa motum habet progressiuum; haud secus, ac videmus multis globis elasticis in linea recta iuxta se dispositis et aequalibus, quorum extremus si percutiatur, non totam globorum seriem, sed solum extremum oppositum moueri. Et hac ratione haud difficulter quisque videt, posse quoque in motu fluidorum eandem quantitatem virium perpetuo conseruari, omnino sicut in motu corporum elasticorum se inuicem percutientium ; imo necessario id fieri ob summam elasticitatem fluidorum corpusculis minimis insitam. Iam vero rem ipsam aggrederer, nisi quibusdam vel solum conseruationis virium viuarum nomen stomachum mouere perspectum haberem. Horum in gratiam monendum duco principium hoc conseruationis virium viuarum minime differre a principio quod primum ab Hugenio usurpatum dein ab omnibus Geometris sine vlla controuersia receptum fuit; nimirum corpora vi grauitatis ad descensum vtcunque sollicitata eam acquirere velocitatem vt si singula rursus velocitate sua finali directe ascendant, vsque ad statum quietis commune centrum grauitatis ad

*Tom.* II.              P                  pri-

priftinam altitudinem redeat ; cui hoc Hugenii principium magis arridet, is eadem facilitate rem expediet, addendum autem eft hoc alterum , velocitates fluidorum per vas inaequaliter amplum fluentium vbique effe amplitudinibus reciproce proportionales : Hifce itaque duobus principiis totum argumentum abfoluemus.

Prop. I. *Problema.* Data celeritate, qua fuperficies aquae in tubo quocunque mouetur, inuenire vim viuam totius maffae aqueae.

*Solutio.* Sit (Fig.I.)vas ABGH, per quod fluat liquor CDFE, cuius fitus proximus fit $cdfe$; habeat fuperficies CD velocitatem, quam acquireret graue cadendo ex altitudine NO. Patet autem, fore velocitatem in LM ad velocitatem fuperficiei CD in ratione reciproca amplitudinum CD et LM vnde fi totum fluidum concipiatur diuifum in ftrata infinita eiusdem altitudinis, quale eft LM$ml$, erit vis viva cuiuslibet ftrati, ficuti ipfius maffa ducta in quadratum velocitatis, id eft, ficuti $\frac{LM}{CD} \times \frac{CD^2}{LM^2}$, feu vt $\frac{CD}{LM}$. Sunt ergo vbique vires viuae in reciproca ratione amplitudinum: Hinc intelligitur, quod facta fuper eodem axe AH alia curua QST tali, vt fit MS vbique aequalis tertiae continue proportionali ad LM et CD fore vim viuam totius maffae aqueae CDFE =fpatio DQTF×NO. Si fymbolis vti velimus, habebimus vim viuam quaefitam$=aav\int\frac{dt}{s}$, vbi $a$ denotat fuperficiem CD, $v$ altitudinem, qua graue cadendo acquirit rit velocitatem iftius fuperficiei ; $dt$ fignificat elementum M$m$, et $s$ amplitudinem vafis in ML. Q. E. I.

*Prop. 2. Theor.* Si tubus (fig. 1.) ABGH verticaliter pofitus fit, atque maffa aquea fuo pondere defcendat in fitum CDFE, quem mox commutet cum fitu *cdfe*; erit differentiale vis viuae feu incrementum vis viuae illo tempufculo acquifitum aequale ei, quam acquirit lapfu per D$d$ cylindrus aqueus cuius bafis eft CD et altitudo DF.

*Dem.* Vis viua acquifita aeftimanda eft ex maffa et altitudine defcenfus: dum vero CDFE peruenit in fitum *cdfe*, idem eft ac fi aqua *cd* FE in fuo loco permanfiffet et aqua CD*dc* in fitum EF*fe* peruenifet ; eft itaque vis viua de nouo acquifita $=$ CD$\times d$D$\times$DF feu, quod perinde eft, CD$\times$DF$\times$D$d$.  Q. E. D.

*Prop. 3. Probl.* Determinare velocitatem aquae qua fluidum fingulis momentis effluit per tubum vtcunque formatum et quocunque foramine perforatum.

*Solutio.* Sit curua quaecunqua BDFG (fig. 2.) cuius applicatae horizontales DC reprefentent refpectiue amplitudines tubi ; Defcenderit aqua ex A in C fitque AC$=t$, CD$=s$ amplitudo foraminis defignata per LM fit$=b$, altitudo tota AM$=c$ ; vis viua totalis infita fluido dum eft in fitu DCMG$=$M: velocitas, quam habet fuperficies DC, aequalis illi quam corpus acquirit cadendo ex alt. $v$; concipiatur nunc aquam ex fitu DCMLGD peruenifse in fitum FEONLGF ; Erit ergo per Prop. Sec. differentiale vis viuae $=$DC$\times$CM$\times$CE$=s\times(c-t)\times dt$ fed poteft idem incrementum aliter fic definiri. Dum aquae superficies esset in DC, erat velocitas in FE$=$ $\frac{s}{s+ds}V u$ (sunt enim velocitates in reciproca ratione amplitudinum) et vis viua aquae FEMLGF erat $= M - svdt$; Iam postquam superficies aquae ex DC peruenit in FE, habet velocitatem $Vu + \frac{du}{2\sqrt{u}}$; vnde cum vires viuae sint in duplicata ratione velocitatum, erit aquae FEMLGF vis insita $=(Ms-vdt)\times(\sqrt{v}+\frac{dv}{2\sqrt{v}})^2 : (\frac{s\sqrt{v}}{s+ds})^2 =$ (neglectis negligendis) $\frac{Msv - ssvvdt + Msdv + 2Mvds}{sv}$, cui quantitati si addatur vis guttulae LMON modo e tubo egressae, habebitur vis, quam habet aqua post situs mutationem; est autem particula aquae LMON$=$DCEF$=sdt$ et habet velocitatem quacum ascendere posset ad altitudinem $\frac{ss}{bb}v$; vnde ipsius vis $= \frac{s^3}{bb}vdt$; ergo vis viua totalis aquae FEONLGF$= \frac{Msv - ssvvdt + Msdv + 2Mvds}{sv} + \frac{s^3}{bb}udt$, a qua proin si auferatur M, habebitur differentiale vis viuae quod adeoque erit $\frac{Msdu + 2Muds - ssuudt}{su} + \frac{s^3 udt}{bb}$. Hinc habetur talis aequatio $s(c-t)dt = \frac{Msdu + 2Muds - ssuudt}{su}$ $+ \frac{s^3 udt}{bb}$. Potest autem per prop. 2. haberi M et datur $s$ per $t$; habetur itaque aequatio inter $t$ et $u$, qua mediante potest determinari velocitas aquae in quocunque situ. Q. E. I.

*Prop.* 4. *Problema.* Determinare velocitates, quibus aqua singulis momentis effluit e tubo cylindrico verticali,

li, fiue ipfius foramen fit finitum fiue infinite paruum.

*Sol.* Sit amplitudo feu fectio cylindri ad axem perpendicularis $=n$, amplitudo foraminis $=1$, altitudo totius cylindri aquei ante effluxum $=c$, altitudo per quam fuprema fuperficies iam defcendit $=t$: Erit ergo vis viua aquae in tubo refiduae (quam fupra vocauimus $M$)$=n(c-t)v$ et quod antea vocauimus generaliter *s* iam conftanter eft *n*; fubftitutis adeoque hifce valoribus in aequatione canonica prodibit

$n(c-t)dt = n(c-t)\,du - nudt + n^3 udt$ ; ponatur $(c-t) = z$ et $nn - 1 = m$ et orietur $-zdz = zdu - mudz$, quae formula ad differentialia logarithmica iuxta methodum paternam reducta atque rite pertractata dat $v = \dfrac{c^{\frac{nn-2}{}}z - z^{\frac{nn-1}{}}}{\frac{nn-2}{(nn-2)c}}$

Q. E. I.

*Coroll.* 1. Si $n = 1$, id eft, fi nullus fit fundus in tubo erit $v = c - z$, ita vt velocitas aquae eadem fit, ac fi graue motu naturaliter accelerato defcendiffet per altitudinem $c - z$; id quod quilibet fine inftituto calculo affequi potuiffet; fuerunt tamen, qui et in hoc cafu crediderunt, aquam effluere eadem velocitate ftatim ab initio, quam corpus acquirit cadendo ex tota altitudine aquae.

*Coroll.* 2. Datur femper locus in tubo, vbi fi peruenerit fuperficies aquea, fit velocitas aquae effluentis maxima; is locus obtinetur, cum fumitur $z = c$: $^{(nn-1)}$ fitque tunc $v = c^{(1:nn-2)}: \overline{nn-2} \times \overline{nn-1}^{1:(nn-2)}$

P 3

$-c$:

$-c : \overline{nn-1} \times \overline{nn-2}^{(nn-1):(nn-2)}$ siue $= c : \overline{nn-1}^{(nn-1):(nn-2)}$ et haec quantitas dat maximam velocitatem, qua superficies aquae in tubo descendere potest; si vero eandem quantitatem multiplicemus per $nn$, habebitur altitudo pro generanda maxima velocitate aquae effluentis quae semper minor est quam $c$; quod si vero $n$ sit infinitum, degenerabit eadem in $c$; vnde per nostram methodum etiam manifestum sit, in casu foraminis infinite parui aquam ea exilire velocitate, qua corpus ascendere possit ad altitudinem aquae. Hicque solus est casus, quem scriptores hydraulicae recte assecuti sunt. Quod si $n$ sit numerus non infinitus, sed tamen sat magnus, erit maxima velocitas aquae effluentis haud multum minor, quam si foramen esset infinite paruum; nam si $n$ fiat $=10$ poterit aqua, maxima sua velocitate effluens, ascendere ad $\frac{97}{100}$ ipsius $c$, hancque velocitatem maximam statim fere a fluxus principio acquiret, nimirum postquam aqua descendit in tubo per spatium $\frac{47}{1047}c$. Haec omnia conueniunt egregie cum experimentis.

*Coroll.* 3. Si $n$ sit numerus magnus, et aqua in tubo iam aliquousque descendere coepit, erunt velocitates quam proxime vti radices altitudinum aquae; quam regulam illi assumserunt, qui de diuisione clepsydrarum egerunt, veluti Cel. Varignon, Mariotte; mihi quoque de clepsydra Sphaerica mari adhibenda aliquando agenti res ita considerata fuit; sed falleret tamen paulisper regula a principio effluxus, nisi $n$ esset numerus admodum magnus. Pro vera diuisione requiritur, vt integretur $\frac{dz}{\sqrt{u}}$ $\frac{dz}{\sqrt{u}}$ id quod semper fieri nequit; series tamen pro hoc negotio dari possunt quibus mediantibus tempus quoque absolutum evacuationis habetur quam proxime in singulis casibus. Idque mihi occasionem dedit, experimenta quam plurima instituendi cum variis tubis, diuersisque foraminibus; et semper tempus depletionis experientia atque calculo idem fuit deprehensum, ita, vt nunc extra dubium posita sit nostra theoria; habeo autem alia experimenta in promtu, quibus res non minus secure ad examen reuocari potest.

*Coroll.* 4. Dederunt allegati Authores curuam pro clepsydra aequabilis descensus, id est, in qua superficies fluidi aequalibus temporibus aequaliter descendit; sed fallit illa et in fine et in principio; atque circa medium satisfacit tantum cum foramen est valde paruum. Vera curua deducitur ex nostra aequatione canonica prop. 3. assumendo pro $v$ constantem et delendo terminum vbi ingreditur $dv$; sed tantum abest quin sit algebraica, vt perueniatur ad differentialia secundi ordinis, si quantitatem M eliminare velimus, eique substituere aequiualentem. Notabile interim est problema circa aequabilem effluxum, quod fit algebraicum: Si nimirum (Fig. 3.) $AND$. curua quaeratur cuius rotatione circa axem BF generetur solidum tale, vt si fluidum in illo descendat, tanquam in clepsydra, aequalibus temporibus aequales quantitates effluant, seu, quod idem est, vt velocitate constante aqua effluat. Sit igitur $FM = t$, $MN = y$, $e =$ altitudini FB, $l =$ altitudini qua graue cadendo acquirit velocitatem aquae effluentis, et erit curua aequabilis effluxus expressa tali aequatione $y^4 t - y^4 l + c^4 l$

$=$

$=o$. habet haec curua multas proprietates, quarum enumeratione, ne nimis sim longus, superfedeo; sed quod minime tacendum puto, est, quod figura haec pro vase aequabilis effluxus eadem est, quae Anglorum cataracta, qua phaenomena aquarum effluentium explicare contenderunt. Caeterum pro maiori confirmatione theoriae nostrae, dicam etiam breuibus de vasis cylindricis, quibus tubi alii cylindrici angustiores annexi sunt, siue verticaliter positi (quorum phaenomena nuper publice exposuit Cel. Bulffingerus) siue horizontaliter: siue abrupti, siue indefinite longi, quorum posteriorum considerationem in rem physiologicam haud parum facere, alia occasione ostendam.

*Prop.* 4. *Probl.* Aqua currente per cylindrum ACDB (fig. 4.) cui tubus cylindricus EF infixus, determinare vbique velocitatem, qua superficies aquae GH descendit.

*Sol.* Potest hic casus tanquam corollarium considerari propositionis tertiae, atque ita determinari vbique velocitas superficiei aqueae, donec tota effluxerit aqua ex vase ACDB; motus autem reliquus per tubum EF per se facile determinatur, quia fit iuxta leges motuum corporum vniformiter acceleratorum. Quod si itaque vas ab initio repletum fuerit vsque in AB dicaturque $AG=t$, $AC=e$, $EF=c$ superficies $GH=n$, amplitudo tubi $EF=1$, altitudo pro generanda velocitate superficiei $GH=v$, mutabitur aequatio canonica factis vbique rite substitutionibus in talem ad casum propositum

fa-

facientem $(c+e-t) \times dt = (e-t+nc) \times du + (nnu-u) \times dt$. Eandem aequationem modo particulari ita erui. Vis viua totius aquae fluentis GCEFEDH est $= n \times (e-t) \times v + nncv$, cuius differentiale est $(ne-nt+nnc)du - nudt$, cui si additur vis viua particulae FP effluxae, quae est $n^3 udt$ habetur totum incrementum vis viuae illo momento generatum, quod per Prop. 2. aequale est $(e-t+c)ndt$, et diuiso vtrobique per $n$, oritur iterum $(c+e-t) \times dt = (e-t+nc) \times du + (nnv-v) \times dt$. Inter has methodos ea differentia est, quod priori res multo generalius expediri potuisset, quam posteriori, ponendo neque vas neque tubum cylindricum sed alia figura praeditum. Pro reducenda posteriori hac aequatione ad quantitates finitas ponatur $t = q+e+nc$ et $v = r - \frac{c}{n+1}$, atque sic illa mutabitur in hanc $-qdq = -qdr + (nn-1)rdq$, quae iterum obseruata debitae constantis additione abit in hanc $q^{1-nn} \times (2r-nnr-q) = (-e-nc)^{1-nn} \times (\frac{ne+e+nc+2c}{n+1})$ vel $\frac{nnr-2r+q}{nn-1} = (nc+2c+ne+e):(-n-1) \times (-e-nc)^{nn-1}$. Et reassumtis quantitatibus $t$ et $v$ oritur $(t-e-nc)^{1-nn} \times (-nnv+2v+c-t+e+\frac{c}{n+1}) = (-e-nc)^{1-nn} \times (\frac{ne+e+nc+2c}{n+1})$ seu $v = (e+c+\frac{c}{n+1}-t):(nn-2) - (e+c+\frac{c}{n+1}) \times (t-e-nc)^{nn-1}:(nn-2) \times (-e-nc)^{nn-1}$.

*Coroll. I.* Si ponatur $c=0$, oritur theorema prop. 4.

*Coroll. II.* Si comparetur velocitas maxima cum velocitate maxima aquae effluentis in cafu $c = o$, inuenietur illa multo maior hac; et quidem differentia eo maior eſt, quo longior eſt tubus EF. Vnde non mirum, ſi tempora exinanitionum eo ſint minora, quo tubi annexi ſunt longiores. Poſſunt autem haec omnia calculo exactius ſubiici tum ratione velocitatum, tum ratione temporum.

*Coroll. III.* Si tubus EF ſit indefinite longus, inuenietur alio ratiocinio ſed non multum abſimili talis aequatio $edu - tdu + nnudt + nntdu - udt = edt - tdt + ntdt$.

*Coroll. IV.* Si tubus EF in F paulo amplior ſit, quam in E tempus depletionis adhuc minus erit; poteſt vero amplitudo in F eo vsque augeri, donec aqua inter effluendum lateribus tubi adhaereſcere deſinat.

*Coroll. V.* Si tubi EF ſint inclinati paucis mutatis idem eſt calculus; quapropter hoſce caſus non attingam excepto illo, quo tubus eſt indefinite longus et horizontalis, quia is in aliis occaſionibus vſui venire poteſt. Quod ſi itaque cylindrus ACDB (fig. 5.) aqua plenus vsque in AB tubum habeat annexum horizontalem DE, indefinite longum, atque aqua in cylindro deſcenderit vsque ad MN, in tubo autem progreſſa fuerit vsque in R, dicatur $AC = a$, amplitudo $AB = n$, foramen $= 1$, $cm = z$ et celeritas aquae MN talis quae acquireretur lapſu ex altitudine $v$, dico aequationem inter $z$ et $v$ fore talem

$$zv - nnzv + nnav - \tfrac{1}{2}zz + \tfrac{1}{2}aa = 0 \text{ vel } v = \frac{aa-zz}{2(z-nnz+nna)}.$$

*Scholium.* Experimenta quoque inftitui cum vafis, cui tubi ftrictiores annexi funt, fed contigit, quod praeuidi, tempora exinanitionum femper maiora effe, quam pro calculo effe deberent; id autem frictionibus tribuendum eft, vbi enim aqua per foramen exilit nulla fere eft frictio; fed cum per canales ftrictiores fluit, obftacula funt latera tuborum. Eft mihi methodus etiam ad calculum reuocandi huiusmodi refiftentias facto vnico experimento; fed omni accuratione inftituendo; neque enim difficulter apparet effe refiftentias in reciproca ratione diametrorum; et directa velocitatum atque tuborum longitudinum hasque refiftentias fubtrahi debere a preffione aquam ad motum follicitante, atque vt fpecimen methodi exhibeam ponam vas aqua repletum *ampliffimum*, cui tubus ftrictus cylindricus horizontaliter infixus, effeque altitudinem aquae fupra tubum$=a$, diametrum tubi $=b$, longitudinemque $=c$, velocitatem aquae exilientis talem quae debeatur altitudini $x$ et erit refiftentia tubi $=\frac{nc\sqrt{x}}{b}$ (per $n$ intelligitur numerus experimento erutus), qua fubtracta a preffione aquae (qnae in hypothefi foraminis refpectu vafis minimi proportionalis eft altitudini aquae fupra tubum) oritur $\frac{nc\sqrt{x}}{b} - a$, quae quantitas fimul exprimit altitudinem ad quam aqua velocitate fua afcendere poffet; vnde $\frac{nc\sqrt{x}}{b} - a = x$; et $x = \frac{nncc - 2abb + nc\sqrt{(nncc - 4abb)}}{2bb}$.

*Prop.* 5. Poteſt alio modo concinnius problema generale propoſitionis tertiae ſolui. Sit nimirum (fig. 6.) BDG curua, cuius axis verticalis eſt AM; abſciſſae MC denotent altitudines aquae reſiduae et applicatae CD exprimant amplitudines vaſis in eodem loco ſeu ſuperficiem aquae: GM repraeſentet fundum vaſis et LM foramen: fiat nunc alia curua SRP ſuper eodem axe talis, vt vbique ſit CR $=$ tertiae continue proportionali ad DC et LM: ſitque MC$=t$; CD$=s$, GM$=c$, LM$=b$, velocitas aquae effluentis LNOM talis, quae generatur lapſu libero corporis ex altitudine $v$, et erit CR$=\frac{bb}{s}$, et (per prop. 1.) vis viua inſita fluido DCMG $=$ ſpatio CRPM$\times v$; ponamus autem ſpat. CRPM$=$N, conſideremusque decreſcente MC decreſcere N, $t$ et $v$: ita habetur incrementum vis viuae aquae in vaſe reſiduae $=-Ndv-vd$N ſeu (ponendo loco $-d$N valorem $-\frac{bbdt}{s}$) $=-Ndv-\frac{bbvdt}{s}$, cui ſi addatur vis viua particulae LNOM, quae eſt $=-svdt$, orietur incrementum totale vis viuae aequandum cum $-stdt$; ergo $-Ndv-\frac{bbvdt}{s}-svdt=-stdt$, vel $dv+\frac{bb+ss}{N}vdt=\frac{stdt}{N}$; ponatur $\frac{bb+ss}{N}dt=d$P et $\frac{stdt}{N}=d$Q, ergo $dv+vd$P$=d$Q, fiat $v=b^{-P}$R (intelligendo per $b$ numerum cuius logarithmus eſt vnitas) et erit $b^{-P}d$R $= d$Q et R$=\int b^P d$Q, et denique $v=b^{-P}\int b^P d$Q. Q. E. I.

Supereſſent plura alia dicenda; veluti de fluidis elaſticis, de fluidis ex vaſe per duo foramina vtcunq; poſita effluentibus (quod vltimum problema difficillimum eſt) aliisque; haec autem nunc ſufficiant. Caeterum ſequitur quoque

ex

ex theoria praefente, vt corollarium, doctrina fluidorum in tubis ofcillantium, quod argumentum Newtonus attigit in theor. 35. lib. 2. princ. Math. Phil. nat. p. 363. edit. 3. vbi motus iste ofcillatorius recte definitur, atque conformiter cum nostris, quae ita egregie confirmantur: neque certe quicquam minimam patitur exceptionem, modo duo principia *conseruationis virium viuarum* et *velocitatum reciproce amplitudinibus proportionalium* concedantur; prius in dubium vocari nequit, si ad frictiones aliasue resistentias extrinsecas non attendatur; qui aliter sentit, totam mechanicam reiicit, quippe conseruatio virium aliis verbis ab omnibus fuit accepta; quod ad alterum principium, quamuis id non ad rigorem verum sit, potest tamen in plerisque vasis fine vllo fcrupulo accipi, nempe omnibus illis, in quibus nullus fit transitus subitaneus: at si v. g. vas acciperetur (Fig. 7.) ABCD perforatum in E, cui quasi faccus quidam adhaereret in O, nemo non videt motum aquae facco O inclusae longe alium fore quam principium istud postularet; neque adeo theoria extendi potest ad huiusmodi vasa valde irregularia; puto autem in hifce casibus nihil certi statui posse.

This page intentionally left blank.

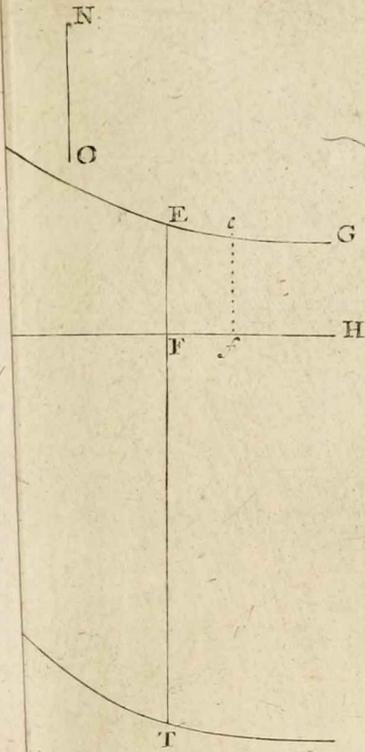
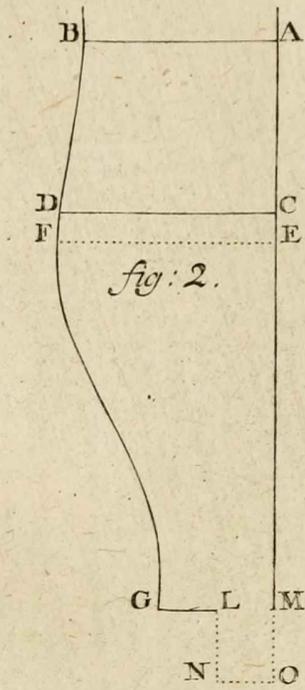

fig: 2.

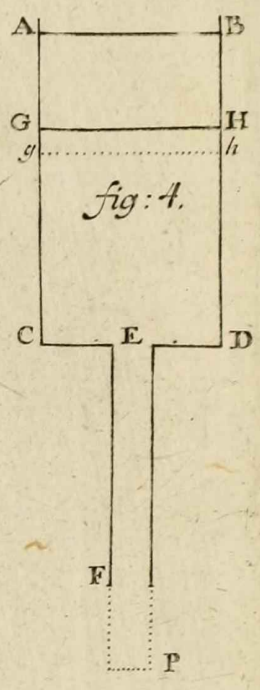

fig: 4.

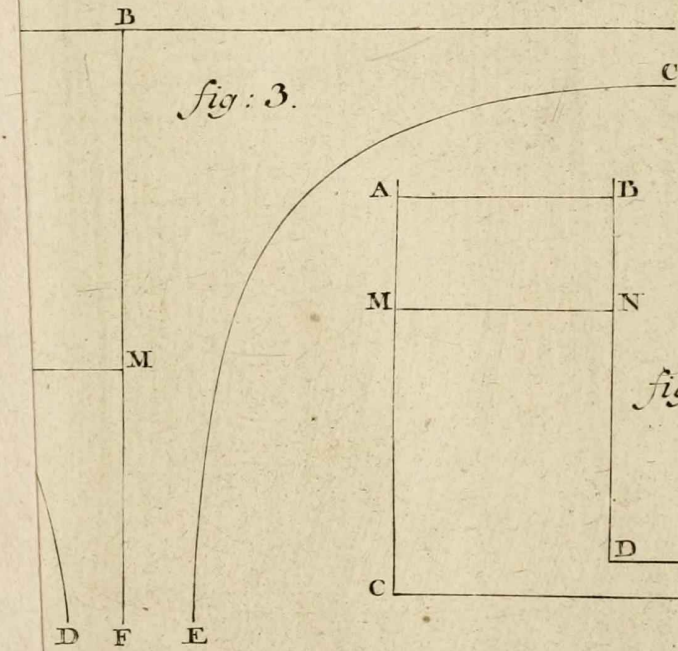

fig: 3.

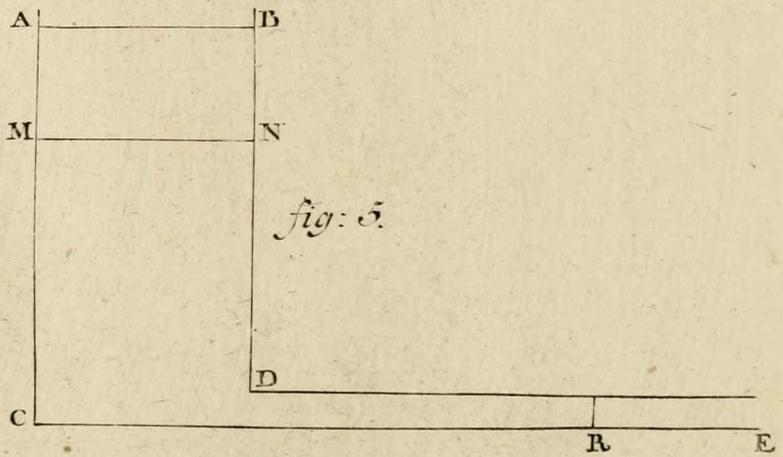

fig: 5.

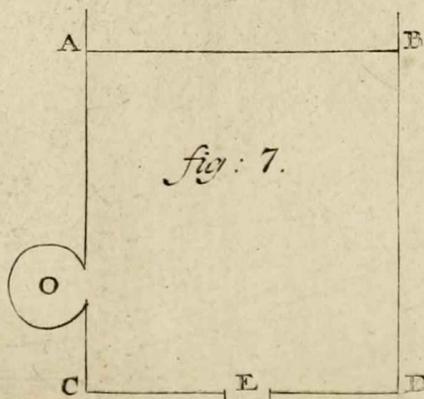

fig: 7.

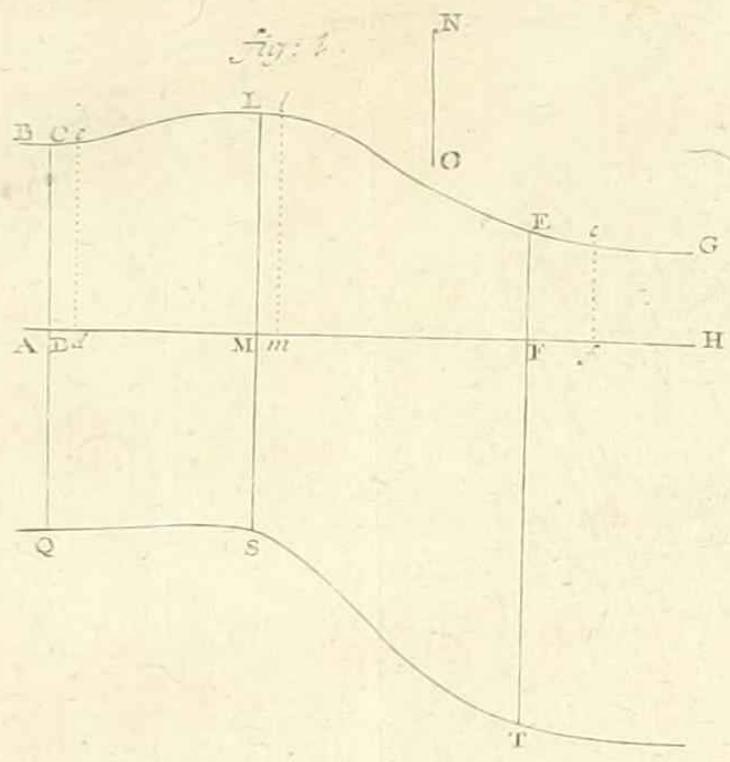
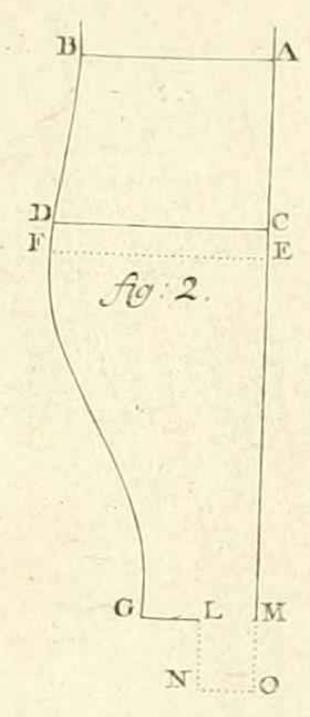
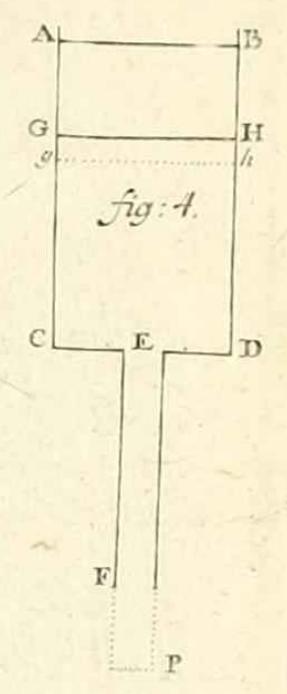
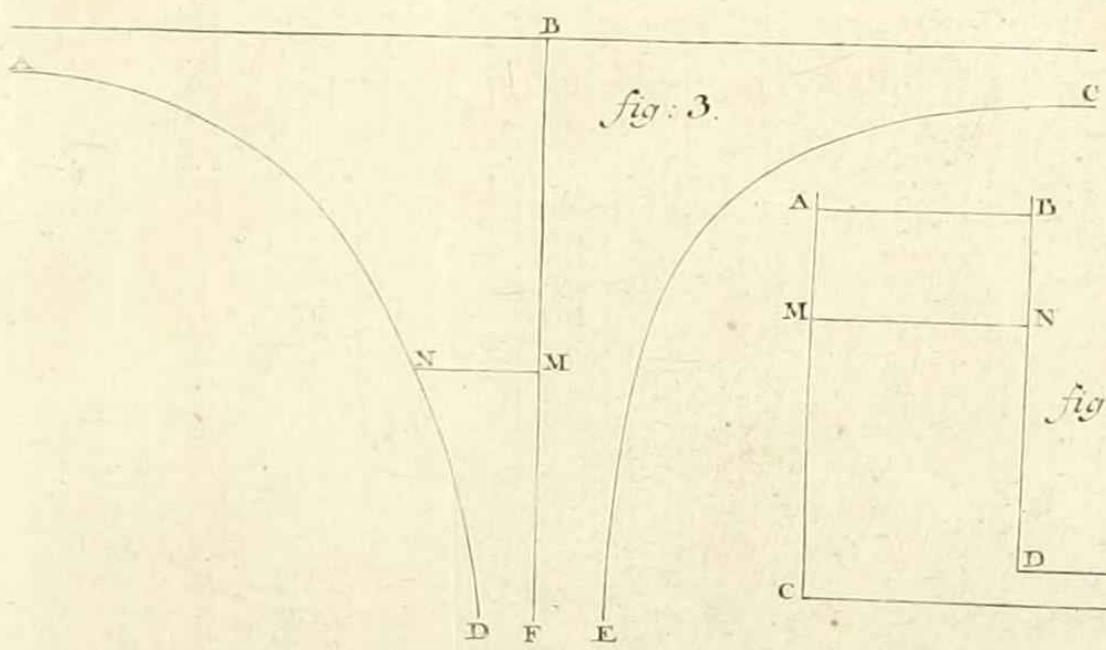
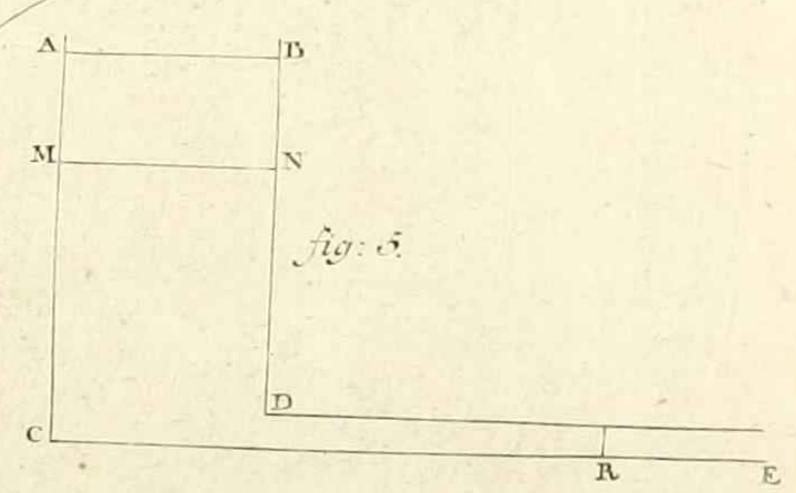
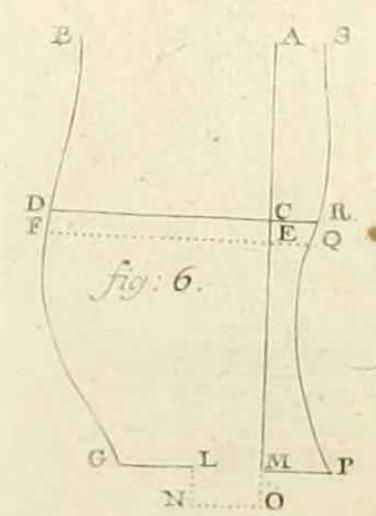
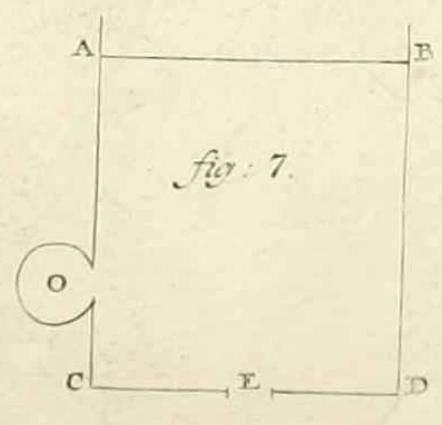